\documentclass[11pt]{article}

\usepackage{tikz}
\usetikzlibrary{matrix,arrows,shapes}
\usepackage{pgf}
\usepackage{url}
  \usepackage{enumerate}
  \usepackage{amsthm}
  \usepackage{amssymb}
  \usepackage{amsmath}
    \usepackage{amscd}
  \usepackage{eucal}
  \usepackage{mathrsfs}
  \usepackage{upgreek}
   \usepackage{dsfont}
      \usepackage{yfonts}
  \usepackage[T1]{fontenc}
  \usepackage{bbm}
  \usepackage{geometry}
\usepackage{array}
\usepackage{booktabs}
\newcommand{\E}{\mathfrak{E}}

\newcommand{\V}{\mathfrak{V}}

\newcommand{\fA}{\mathfrak{A}}

\newcommand{\D}{\mathfrak{D}}

\newcommand{\F}{\mathfrak{F}}

\newcommand{\BV}{\mathfrak{BV}}

\newcommand{\Dcal}{\mathcal{D}}

\newcommand{\Ocal}{\mathcal{O}}
\newcommand{\Scal}{\mathcal{S}}
\newcommand{\Rcal}{\mathcal{R}}
\newcommand{\Tcal}{\mathcal{T}}

\newcommand{\Ci}{\mathcal{C}^\infty} 
\newcommand{\Nat}{\mathrm{Nat}}

\newcommand{\Loc}{\mathrm{\mathbf{Loc}}}       
\newcommand{\Vect}{\mathrm{\mathbf{Vec}}}       

\newcommand{\id}{\mathrm{id}}               
\newcommand{\supp}{\mathrm{supp}}      
\newcommand{\im}{\mathrm{Im}}             
\newcommand{\loc}{\mathrm{loc}}

\newcommand{\reg}{\mathrm{reg}}
\newcommand{\ren}{\mathrm{r}}

\newcommand{\pg}{\mathrm{pg}}

\newcommand{\af}{\mathrm{af}}
\newcommand{\ta}{\mathrm{ta}}
\newcommand{\gh}{\mathrm{gh}}

\newcommand{\mc}{\mathrm{mc}}

\newcommand{\CC}{\mathbb{C}}           
\newcommand{\M}{\mathbb{M}} 	     
\newcommand{\al}{\alpha}

\newcommand{\Ga}{\Gamma}
\newcommand{\de}{\delta}
\newcommand{\De}{\Delta}

\newcommand{\la}{\lambda}
\newcommand{\La}{\Lambda}

\newcommand{\ph}{\varphi}
\newcommand{\T}{\cdot_{{}^\Tcal}}
\newcommand{\TL}{\cdot_{{}^{\Tcal_\Lambda}}}
\newcommand{\TR}{\cdot_{{}^{\TTR}}}
\newcommand{\delT}{\delta^{\sst{\TT}}_{S}}
\newcommand{\delTR}{\delta^{\sst{\TTR}}_S}
\newcommand{\deL}{\delta^{{\Lambda}}_S}
\newcommand{\TT}{\Tcal}
\newcommand{\TTR}{\Tcal_\ren}
\newcommand{\TRH}{\cdot_{{}^{\TTR}}}

\newcommand{\TTL}{\Tcal_{\!\sst{\Lambda}}}

\newcommand{\DDp}{\Gamma'_{\Delta_{D}}}
\newcommand{\DD}{\Gamma_{\Delta_{D}}}
\newcommand{\DC}{\Gamma_{\Delta}}

\newcommand{\HL}{\Gamma_{\Lambda}}
\newcommand{\paqft}{{p\textsc{aqft}}}

\newcommand{\eom}{{\textsc{eom}}}
\newcommand{\qme}{{\textsc{qme}}}

\newcommand{\cme}{{\textsc{cme}}}
\newcommand{\mwi}{{\textsc{mwi}}}
\newcommand{\sst}[1]{\scriptscriptstyle{#1}}  
\newcommand{\minus}{\sst{-1}}   
\newcommand{\pa}{\partial}                              
\newcommand{\be}{\begin{equation}}
\newcommand{\ee}{\end{equation}}

\newcommand{\Lap}{\bigtriangleup}

\newcommand{\os}{\stackrel{\mathrm{o.s.}}{=}}

 \author{\null\\ Klaus Fredenhagen, Katarzyna Rejzner \\
  \null\\
  \null\\
        \small{ II. Inst. f. Theoretische Physik, Universit\"at Hamburg,}\\
    \small{Luruper Chaussee 149, D-22761 Hamburg, Germany}\\ 
\small{\texttt{klaus.fredenhagen@desy.de,katarzyna.rejzner@desy.de}}}

\title{Batalin-Vilkovisky formalism in perturbative algebraic quantum field theory}

\begin{document}
\date{}
 \maketitle

  \theoremstyle{plain}
  \newtheorem{definition}{Definition}[section]
  \newtheorem{theorem}[definition]{Theorem}
  \newtheorem{proposition}[definition]{Proposition}
  \newtheorem{corollary}[definition]{Corollary}
  \newtheorem{lemma}[definition]{Lemma}
  
  \theoremstyle{plain}
  \newtheorem*{Main}{Main Theorem}
  \newtheorem*{MainT}{Main Technical Theorem}

  \theoremstyle{definition}
  \newtheorem{remark}[definition]{Remark}

 \theoremstyle{definition}
  \newtheorem{ass}{\underline{\textit{Assumption}}}[section]

\begin{abstract}
On the basis of a thorough discussion of the Batalin-Vilkovisky formalism for classical field theory presented in our previous publication, we construct in this paper the Batalin-Vilkovisky complex in perturbatively renormalized quantum field theory. The crucial technical ingredient is an extended notion of the renormalized time-ordered product as a binary product equivalent to the pointwise product of classical field theory. Originally, in causal perturbation theory, the time-ordered product is understood merely as a sequence of multilinear maps on the space of local functionals. Our extended notion of the renormalized time-ordered product (denoted by $\TR$) is consistent with the old one and we found a subspace of the quantum algebra which is closed with respect to $\TR$. On this space
the renormalized Batalin-Vilkovisky algebra is then the classical algebra but written in terms of the time-ordered product, together with an operator which replaces the ill defined graded Laplacian of the unrenormalized theory. We identify it with the anomaly term of the anomalous Master Ward Identity of Brennecke and D\"utsch. Contrary to other approaches we do not refer to the path integral formalism and do not need to use regularizations  in intermediate steps.
\end{abstract}
\section{Introduction}
\label{intro}
A powerful method for the treatment of quantum field theories with gauge symmetries is the Batalin-Vilkovisky formalism which extends the BRST method \cite{BRST1,BRST2,CR} and allows to discuss these theories without reference to a specific gauge fixing. Its main advantage is the simultaneous treatment of equations of motion and gauge symmetries in terms of homological algebra.

Its application to relevant physical theories is, however, somewhat formal, since the mathematical methods are designed for finite dimensional situations (see for example \cite{Froe}) whereas the examples from physics are typically infinite dimensional. Moreover, the  formulation of the so-called Quantum Master Equation ({\qme}) which is used as the starting point for the construction of a renormalized quantum field theory, suffers from the occurrence of ill defined terms. 

The problem to incorporate the renormalization into the BV formalism is present since the first papers of Batalin and Vilkovisky \cite{Batalin:1981jr,Batalin:1983wj,Batalin:1983jr}. In \cite{Batalin:1983jr} the authors comment on this problem pointing out the existence of divergences and they propose to deal with them by applying some regularization scheme which puts the divergent terms of the {\qme} at 0. In \cite{Troost} it was proposed to use instead a regularization that gives to these terms finite non-zero values. This approach allowed to analyze the anomalies in a more systematic way and relate them to obstructions in fulfilling the \textsc{qme}. The regularization used in \cite{Troost} is the Pauli-Villars scheme and the discussion is restricted only to the 1-loop order. A method valid for higher loop orders was proposed in \cite{Pa95}, but the regularization  scheme used there is non-local. The dimensional regularization and renormalization in the context of BV formalism were discussed in \cite{Tonin}. The BPHZ renormalization is discussed in \cite{JPT}. All of the mentioned approaches rely on some regularization scheme and involve arbitrary choices. From the conceptual point of view it is still unclear how the \textsc{qme} should be interpreted in the renormalized theory. An alternative treatment of the \textsc{qme} which involves certain extension of the field-antifield formalism was presented in \cite{Barnich}.

An approach to a rigorous formulation of the {\qme} has recently been performed by Costello \cite{Costello}. He replaces the Quantum Master Equation by a  family of regularized equations which are interpreted in terms of different scales. An unsatisfactory aspect of this approach (which is shared by many regularization schemes in quantum field theory) is that the problem which one wants to solve cannot be precisely formulated a priori.

Many rigorous approaches to quantum field theory are based on the euclidean version of the theory where spacetime is replaced by a Riemannian space. This makes the path integral more reliable and simplifies the analysis of singularities. Moreover, concrete calculations often give the same results, independent of the signature of the spacetime metric. But the Osterwalder-Schrader theorem \cite{OS} on which the transition between euclidean and Lorentzian structures is based holds only under certain conditions which are not generally valid for pseudo-Riemannian manifolds. Moreover, some of the crucial properties of quantum field theory, in particular the local commutativity of mutually spacelike localized observables, are not directly visible in the euclidean version. As a consequence, the fact that the dynamics within a globally hyperbolic subregion is completely independent from the dynamics outside of this region\footnote{For quantum field theory this was first proved in \cite{BF0}.} has no counter part in the euclidean theory. 
We therefore prefer to work directly on Lorentzian spacetimes.

The path integral can be understood as a linear functional on the space of functionals of field configurations. This functional contains in principle the information on the dynamics as well as on the state. But whereas the dynamics is locally determined the state necessarily involves global information. It is therefore desirable to disentangle these two aspects and to separate the dynamics from the specification of the state. Actually, this is the aim of Algebraic Quantum Field Theory as introduced by Haag et al long ago \cite{Haag0,HK}, and on the basis of causal perturbation theory, as proposed by Stueckelberg \cite{Stuck} and Bogoliubov \cite{BS} and rigorously developed by Epstein and Glaser \cite{EG}, a corresponding disentanglement is possible also for renormalized perturbative quantum field theory \cite{BF0}.

The basic idea is to construct inductively the time ordered product as a sequence of symmetric multilinear maps $\TT_n$ of $n$ local functionals of field configurations into the operator algebra of the quantum theory. This construction is (up to finite renormalizations) fixed by the requirement that
\be\label{causalfactorization}
\TT_n(F_1,\dots,F_n)=\TT_k(F_1,\dots,F_k)\star \TT_{n-k}(F_{k+1},\dots,F_n)
\ee
holds whenever there is a Cauchy surface $\Sigma$ such that the functionals $F_1,\dots F_k$ are localized in the future of $\Sigma$ and the functionals $F_{k+1},\dots,F_n$ in the past ({\it causal factorization}). Here $\star$ denotes the operator product of the quantum theory.  

It was shown in \cite{BDF} that the unrenormalized time ordered product $\T$ is equivalent to the pointwise product of functionals. The equivalence is induced by an invertible linear operator $\TT$, called the time ordering operator, which formally coincides with the convolution with the "Gaussian measure" associated to the path integral for the free theory, and was first used in the flow equation approach to renormalization in the line of Polchinski \cite{Salmhofer2}. In the non-renormalized theory it is important that we have an algebraic structure with two products: $\T$ and $\star$ and the relation between them is provided by the causal structure of the spacetime.
A question left open in \cite{BDF} was whether also the renormalized time ordered product can be extended to a binary product on a suitable space of functionals. In this paper we prove that this is indeed the case. The arising product $\TR$ turns out to be equivalent to the pointwise product and is therefore in particular commutative and associative. Whereas commutativity is a direct consequence of the symmetry requirement for time ordered products $\TT_n$, associativity could not be checked before since the domain of these maps is not invariant. 

Having the renormalized extended time ordered product and the operator $\TTR$ inducing the equivalence with the pointwise product of classical field theory at our disposal we can now transport the structure of classical BV theory into quantum field theory. The classical BV theory was revisited in our previous paper \cite{FR} where special emphasis was put on the algebraic and differential geometric aspects.  It turns out that important structures of the quantum BV formalism can be described on the algebraic level, with the use of the two products we have in the quantum algebra: $\TR$ and $\star$.

The crucial observation is now that, under  $\TTR$, the identities which hold for local functionals of the field in classical physics remain no longer valid in quantum physics.
The reason is that the ideal characterizing the dynamics is generated from the field equation by the operator product $\star$, not by the time ordered product $\TR$.
 This phenomenon was already investigated by Brennecke and D\"utsch in their seminal paper \cite{BreDue} without relating it to the BV formalism. These authors found that the violation of the {\mwi} \cite{DF02} was a local functional and termed the relation "anomalous Master Ward Identity" \footnote{There is an obvious analogy to the Quantum Action Principle \cite{PSor}. See \cite{BreDue} for details.}. The relevance of this relation for a proper formulation of the BV formalism in perturbative algebraic quantum field theory was first recognized by Hollands in his paper on the renormalization of Yang Mills theories on curved spacetimes \cite{H}. In our paper we show that it indeed induces a renormalized version of the Quantum Master Equation.  

\section{Nonrenormalized time-ordered products}\label{general}
\subsection{Scalar field}\label{scalar}
We start with the simple example of the free minimally coupled scalar field on a globally hyperbolic spacetime $M$. The configuration space of the theory is the space of smooth functions $\E(M)=\Ci(M)$. The observables of the theory are the smooth functions on this space\footnote{Smoothness has to be understood in the sense of calculus on locally convex vector spaces. See \cite{Neeb,Ham,Michor} for a review.}. Among  them an important role is played by the local ones, i.e. those which are of the form
\be
F(\ph)=\int dx f(j_x(\ph))
\ee
with a smooth function $f$ on the jet bundle, where $j_x(\ph)=(x,\ph(x),\pa\ph(x),\dots)$ is the jet of $\ph$ at $x$, and a volume form $dx$ which may be chosen in our case as the volume form associated to the Lorentzian metric. These functionals have functional derivatives with support on the thin diagonal
\be
\supp F^{(n)}(\ph)\subset\{(x_1,\dots,x_n)\in M^n|x_1=\dots =x_n\}
\ee
and their wave front sets are orthogonal to the tangent bundle of the thin diagonal, considered as a subset of the tangent bundle of $M^n$. Let $\F_\loc(M)$ denote the space of local and $\F(M)$ of multilocal functionals (products of local ones). 
Both $\F_\loc$ and $\F$ are covariant functors from the category of globally hyperbolic spacetimes $\Loc$ with causal isometric embeddings as morphisms to the category $\Vect$ of locally convex vector spaces. 
For details concerning the formulation in the language of category theory, see \cite{BFV}. 

In the next step we introduce the dynamics by means of an action functional $S$. Since neither our spacetimes nor the support of typical configurations are compact we cannot identify $S$ with a function on $\E(M)$. Instead we follow \cite{BDF} and define a generalized Lagrangian $L$ as a natural transformation between the functor of test function spaces $\D:\Loc\rightarrow \Vect$ and the functor $\F_\loc$ such that it satisfies 
\be\label{L:supp}
\supp(L_M(f))\subseteq \supp(f)\,,
\ee
and the additivity rule 
\be\label{L:add}
L_M(f+g+h)=L_M(f+g)-L_M(g)+L_M(g+h)\,,
\ee
for $f,g,h\in\D(M)$ and $\supp\,f\cap\supp\,h=\emptyset$.  
The action $S(L)$ is now defined as an equivalence class of Lagrangians  \cite{BDF}, where two Lagrangians $L_1,L_2$ are called equivalent $L_1\sim L_2$  if
\be\label{equ}
\supp (L_{1,M}-L_{2,M})(f)\subset\supp\, df\,, 
\ee
for all spacetimes $M$ and all $f\in\D(M)$. 

In order to avoid ill defined terms we consider for the time being only regular functionals $F\in\F_{\reg}(M)$. Here we call a map $F:\E(M)\to\CC$ regular whenever it is smooth and all its functional derivatives are smooth densities with compact support. The regular functionals form a Poisson algebra with the pointwise product 
\be
m:\left\{\begin{array}{ccc}\F_{\reg}(M)\otimes\F_{\reg}(M) & \to &\F_{\reg}(M)\\
                                                  F\otimes  G & \mapsto & F\cdot G
                                                  \end{array}
                                                  \right. 
\ee
where $(F\cdot G)(\ph)=F(\ph)G(\ph)$, and with the Peierls bracket as the Poisson bracket,
\be
[F,G]=\langle F^{(1)},\Delta G^{(1)}\rangle \ .
\ee
Here $\De=\De_A-\De_R$ where $\De_{A,R}$ are the advanced and retarded, respectively, propagators of the Klein Gordon equation, considered as maps from smooth compactly supported densities to smooth functions. 

The observables of the quantized theory are constructed as formal power series in $\hbar$ with coefficients in the space of functionals on $\E(M)$.
Let $\fA_{\reg}(M)=\F_{\reg}(M)[[\hbar]]$ denote the space of regular quantum observables. On $\fA_{\reg}(M)$ we define two products. The first product is the operator product
\be\label{star product}
A\star B\doteq m\circ \exp({i\hbar \DC})(A\otimes B) \ ,
\ee
where  $\DC$ is the functional differential operator
\be\label{star product2}
\DC\doteq\frac{1}{2}
                  \int \De(x,y)\frac{\de}{\de\varphi(x)}\otimes\frac{\de}{\de\varphi(y)}\,.
\ee
The complex conjugation satisfies the relation $\overline{F\star G}=\overline{G}\star\overline{F}$,
therefore we can use it to define an involution  $F^*(\ph)\doteq\overline{F(\ph)}$. The resulting structure is a $*$-algebra $(\F_{\reg}(M)[[\hbar]],\star)$ which may be understood as a deformation quantization of the Poisson algebra of classical field theory \cite{BF0,DF}.

The second product is the time ordered product
\be
A\T B=m\circ \exp(i\hbar \DD')(A\otimes B)\,,
\ee
with the functional differential operator
\be\label{}
\DD'\doteq\int \De_D(x,y)\frac{\de}{\de\varphi(x)}\otimes\frac{\de}{\de\varphi(y)}\,,
\ee
where $\De_D=\frac12(\De_A+\De_R)$ is the Dirac propagator. Due to the support properties of the propagators, it coincides for functionals with time ordered supports with the operator product. Moreover, it is equivalend to the pointwise product of classical field theory by the linear operator 
\[
\TT(F)\doteq e^{i\hbar\DD}(F)\,,
\]
with 
\be\label{}
\DD\doteq\int \De_D(x,y)\frac{\de^2}{\de\varphi(x)\de\varphi(y)}\,,
\ee
i.e.
\be\label{Tproduct}
F\T G\doteq \Tcal(\Tcal^{\minus}\cdot\Tcal^{\minus}G) \ .
\ee
$\TT$ is the time ordering operator mentioned in the introduction. It is invertible and its inverse is obtained by replacing $\DD$ by $-\DD$. 

The time ordered product provides us with means to introduce the interaction using the local S-matrices. For an interaction $V\in\fA_\reg(M)$ the formal S-matrix is defined as the time-ordered exponential:
\be\label{Smatrix}
\Scal(V)\doteq e_{\sst{\TT}}^V=\TT(e^{\TT^{\minus}V})\,.
\ee
We can now define the relative S-matrix for $V,F\in\fA_\reg(M)$ by the formula of Bogoliubov:
\be\label{Bog}
\Scal_V(F)\doteq\Scal(V)^{\star-1}\star \Scal(V+F)\,.
\ee
Interacting quantum fields are generated by $\Scal_{iV/\hbar}(F)$ and we can write them as formal power series:
\be\label{Rv}
\frac{d}{d\lambda}\Big|_{\lambda=0}\Scal_{iV/\hbar}(\lambda F)=\sum\limits_{n=0}^\infty\frac{1}{n!} R_{n,1}(V^{\otimes n},F)\equiv R_V(F)\,,
\ee
where the maps $R_{n,1}$ are called retarded products. More explicitly the intertwining map $R_V$ can be written as
\be\label{RV}
R_V(F)=\left(e_{\sst{\TT}}^{iV/\hbar}\right)^{\star\minus}\star\left(e_{\sst{\TT}}^{iV/\hbar}\T F\right)\,.
\ee
When we switch on the interaction, also the star product has to change. A natural definition can be obtained with the use of the intertwining map $R_V$. We define the interacting star product as:
\be\label{interacting:star}
F\star_V G\doteq R_V^{\minus}\left( R_V(F)\star R_V(G)\right)\,,
\ee
where the inverse of $R_V$ is given by:
\[
R_V^{\minus}(F)=e_{\sst{\TT}}^{-iV/\hbar}\T\left(e_{\sst{\TT}}^{iV/\hbar}\star F\right)\,.
\]

In order to perform the construction of the BV complex along the lines of \cite{FR} we need to extend the algebra of functionals with its derivations (the {\it antifields}), i.e. vector fields. We identify them with smooth maps $X$ from $\E(M)$ to itself. We restrict ourselves to maps that have their image in  $\E_c(M)$ (compactly supported sections). The associated derivation is 
\be
(\partial_XF)(\ph)=\langle F^{(1)}(\ph),X(\ph)\rangle \ .
\ee
The spacetime support of a vector field $X$ is defined in the following way:
\be\label{suppder}
\begin{split}
\supp\, X=\{x\in M|\forall \text{ neigh. }U\text{ of }x\ & \exists F\in\F(M), \supp\,F\subset U\ \text{ such that }\partial_XF\neq 0 \\
\text{or } \exists\ \ph,\psi\in\E(M),\supp\,\psi\subset U & \text{ such that }X(\ph+\psi)\neq X(\ph)\}\ .
\end{split}
\ee
We define $\V_\reg(M)$ to be the space of smooth, compactly supported vector fields with image in  $\E_c(M)$ such that all functional derivatives are smooth densities. We call such vector fields regular. 
It was already discussed in \cite{FR} that one can define on $\V_\reg(M)$ the Koszul map $\delta_S$,
\be
\delta_S(X)=\partial_XL(f), \ f\equiv1 \text{ on }\supp X \ .
\ee  
Here $L$ is the generalized Lagrangian and $S$ is the associated action. The image of $\delta_S$ is the ideal in $\F_{\reg}(M)$ generated by the field equation. The space $\Lambda\V_\reg(M)$ of alternating vector fields, equipped with the Koszul map as a differential and the Schouten bracket $\{\cdot,\cdot\}$ as an odd graded Poisson bracket (the {\em antibracket}) is then the BV complex for the classical scalar field.

In analogy to the finite dimensional case, vector fields on $\E(M)$ can be seen from two viewpoints: on one hand as derivations of $\F(M)$ and on the other hand as sections of the tangent bundle, i.e. maps from $\E(M)$ to $\E_c(M)$. These two roles played by vector fields have their consequences for the definition of the time ordering operator on $\V_\reg(M)$. Indeed, if we think of an element $X\in \V_\reg(M)$ as a section, then $\TT$ acts on it simply as a differential operator and we can put forth a following definition:
\be\label{timeordvf}
(\TT X)(\ph)(x)\doteq(\TT X_x)(\ph)\,,
\ee
where $X_x(\ph)=X(\ph)(x)$.

The transformation of the associated derivation is now determined by the principle
to use $\TT$ as a mean to transport the classical structure to the quantum algebra. In this spirit we can associate with $Y\in\TT(\V_\reg(M))$ an operator  on $\TT(\F_\reg(M))$ defined as
\be\label{timeorder}
\partial^{\sst{\TT}}_Y F=\mathcal{T}\langle \mathcal{T}^{-1}Y,\mathcal{T}^{-1}F^{(1)} \rangle\,\qquad F\in\TT(\F_\reg(M))\,.
\ee
From the above formula it is evident that $\partial^{\sst{\TT}}_Y$ is a derivation of $\TT(\F_\reg(M))$ with respect to the time ordered product $\T$:
\be
\partial^{\sst{\TT}}_Y(F\T G)=(\partial^{\sst{\TT}}_Y F)\T G+F\T(\partial^{\sst{\TT}}_Y G)\,,
\ee
Moreover we obtain the following identity:
\be
\partial^{\sst{\TT}}_{\mathcal{T}X}=\mathcal{T}\circ\partial_X\circ\mathcal{T}^{-1}
\ee
The construction we performed shows that we can recover in a natural way all the classical structures of the BV complex in the quantum algebra, but they are defined with respect to the time-ordered product, not with respect to the operator product. 
Since $\T$ is still a graded commutative product (in contrast to $\star$), the BV complex can be defined. 

The graded algebra of antifields is transformed into $\TT(\Lambda\V_\reg(M))$. This algebra is equipped with the time ordered Schouten bracket $\{.,.\}_{\TT}$ defined as:
\be
\{X,Y\}_{\sst{\TT}}=\mathcal{T}\{\mathcal{T}^{-1}X,\mathcal{T}^{-1}Y\}\,.
\ee

Now we want to see how the ideal generated by the equations of motion is transforming under the time ordering. We identify it as the image of the time-ordered Koszul operator:
\be
\delT=\mathcal{T}\circ\delta_{\mathcal{T}^{\minus}S}\circ\mathcal{T}^{-1}\,,
\ee
where $S\in\TT(\F_\reg(M))$. Before characterizing the quantum ideal generated by the equations of motion, we need one more definition. We already defined the time ordered product  $\T$ of antifields, but we need also the operator product $\star$. The definition is quite natural if we treat vector fields as functions $\E(M)\rightarrow\E_c(M)$ and apply to them the operator $\exp({i\hbar \DC})$ defined by (\ref{star product}) and (\ref{star product2}). 

Let us now have a closer look at the image of $\delT$. Acting  on a time-ordered vector field $X\in\TT(\V_\reg(M))$ with $\delT$ we obtain
\[
\delT(X)=\TT(\delta_{\mathcal{T}^{\minus}\!S}(\TT^{\minus} X))=m \circ e^{i\hbar\DDp}\left(\int\!\, X_x\otimes \frac{\delta S}{\delta \ph(x)}\right)\,,
\]
where in the second step we used the Leibniz rule. Since $S$ is a functional of second order in $\ph$, the expansion of $e^{i\hbar\DD}$ has only two nontrivial terms and we finally obtain:
\be
\delT(X)=\delta_S(X) +i\hbar\Lap\! X\label{QMO}\,,
\ee
where $\Lap$ is a map that acts on regular vector fields $\V_\reg(M)$ like a divergence\footnote{This operator is in the literature denoted by $\Delta$, but we use here a slightly different symbol $\Lap$, to distinguish it from the causal propagator $\Delta(x,y)$.}:
\[
\Lap X\doteq \int\! \frac{\delta X_x}{\delta\ph(x)},\qquad X\in\V_\reg(M)\,.
\]
This operator can be extended also to multi-vector fields $\La\V_\reg(M)$ in such a way that it becomes a differential, i.e. $\Lap^2=0$ is fulfilled. Explicitly we can write $\Lap$ as:
\[
\Lap Q=(-1)^{(1+|Q|)}\int \frac{\delta^2 Q}{\delta\ph^\ddagger(x)\delta\ph(x)},\qquad Q\in\Lambda\V_\reg(M)\,.
\]
where we formally identified the antifields $\ph^\ddagger$ with the functional derivatives $\frac{\delta}{\delta\ph}$, such that a vector field $X$ can be written in the form $X=\int X_x\ph^{\ddagger}(x)$.

The operator $\Lap$ has also some nice properties with relation to the antibracket. For example it holds:
\be\label{Delta:bracket}
\{P,Q\}= \Lap(PQ)-\Lap(P)Q-(-1)^{|P|}P\Lap\!(Q)\,,
\ee
where $P,Q\in\La\V_\reg(M)$.
Moreover, using (\ref{Delta:bracket}) and the nilpotency of $\Lap$, one can show that:
\be\label{Delta:bracket2}
\Lap\{P,Q\}=-\{\Lap(P),Q\}-(-1)^{|P|}\{P,\Lap\!(Q)\}\,.
\ee
The graded algebra $\La\V_\reg(M)$ together with the antibracket $\{.,.\}$ and the differential $\Lap$ form a structure, which is called in mathematics the BV-algebra\index{BV!algebra}.

Note that since the time ordering commutes with both derivatives $\frac{\delta}{\delta\ph(x)}$ and $\frac{\delta}{\delta\ph^\ddagger(x)}$, it also commutes with $\Lap$. Hence we obtain
\be\label{Delta:Tbracket}
\{X,Y\}_{\TT}= \Lap(X\T Y)-\Lap(X)\T Y-(-1)^{|X|}X\T \Lap(Y)\,,
\ee
where $ X,Y\in\TT(\La\V_\reg(M))$.
Now we can come back to the problem of comparing the quantum and the classical ideal of {\eom}'s. To see the relation between them, we use the fact that 
\be\label{identity:S:star}
\int\,X_x\cdot\frac{\delta S}{\delta \ph(x)}=\int\, X_x\star\frac{\delta S}{\delta \ph(x)}\,,
\ee
 and we can rewrite (\ref{QMO}) as:
\be\label{tkoszul}
\delT(X)=\int\,X_x\star\frac{\delta S}{\delta \ph(x)}+i\hbar\! \Lap\!(X)\,.
\ee
In this formula both the time-ordered and the $\star$-product appear 
and it is natural to ask, if there is a $\star$-transformed version for the antibracket. In analogy to (\ref{Delta:bracket}) and (\ref{Delta:Tbracket}) we can define it as\footnote{Note that this is \textit{not} a Poisson bracket, essentially because $\star$ is not graded commutative. Nevertheless $\{.,Y\}$ defines a derivation with respect to $\star$ if $\frac{\delta Y}{\delta\ph(x)}$ is central.}:
\be\label{Delta:star:bracket}
\{X,Y\}_{\star}= \Lap(X\star Y)-\Lap(X)\star Y-(-1)^{|X|}X\star \Lap(Y)\,.
\ee
This can also be written as:
\be\label{antibracketstar}
\{X,Y\}_{\star}=-\int\!\left(\!\frac{\delta X}{\delta\ph(x)}\star\frac{\delta Y}{\delta\ph^\ddagger(x)}+(-1)^{|X|}\frac{\delta X}{\delta\ph^\ddagger(x)}\star\frac{\delta Y}{\delta\ph(x)}\!\right)\,,
\ee
In this new notation we can write (\ref{tkoszul}) as:
\be\label{tkoszul2}
i\hbar\! \Lap\!(X)=\{X,S\}_{\TT}-\{X,S\}_\star\,.
\ee
According to this  we can interpret $\Lap$ as an operator describing the difference between the classical ideal of equations of motion represented by the image of $\{.,S\}_\TT$ and the quantum one, characterized as the image of  $\{.,S\}_\star$.
Using the identity (\ref{identity:S:star}) it is easy to see that the operator $\{.,S\}_\star$ is a derivation with respect to the $\star$-product. We can view it as the quantum Koszul map of the free action. 
The fact that $\{.,S\}_\TT$  and $\{.,S\}_\star$ differ  by a $\hbar$-order term corresponds to the Schwinger-Dyson type equations. The operator $\{.,S\}_\star$ is not a derivation with respect to the time-ordered product, but using (\ref{tkoszul2}) and (\ref{Delta:Tbracket}) we can see that it holds:
\be\label{derivation:T}
\{X\T Y,S\}_\star-\{X,S\}_\star\T Y-(-1)^{|X|}X\T \{Y,S\}_\star=-i\hbar\{X,Y\}_{\TT}\,.
\ee
\subsection{Quantum master equation and the quantum BV operator}\label{algQFT}
In the previous section we considered only the example of a scalar field, but
the really interesting story in the BV quantization starts when the action has symmetries. 
Here symmetries are defined as vector fields $X\in\V(M)$ with $\partial_XS=0$. Usually, one divides out the trivial symmetries. i.e. those which vanish on solutions, and chooses a subspace of representatives of equivalence classes in the space of symmetries. This subspace, however, might not be closed under the Lie bracket (case of open algebras). In order to avoid this complication one can work with the space of all symmetries, but then the cohomological problem has to be stated differently and one uses the tools of homological perturbation theory. Since we don't want to introduce too many technical details here, we present our formalism for the case of closed algebras. This is justified, since many interesting physical examples like Yang-Mills theory and general relativity fall into this class. 

Let us now review the BV formalism in the classical theory.
In the first step one constructs the space of alternating multilinear forms (the so-called ghosts) on the space of symmetries  with values in the functions on the configuration space. It is a graded algebra and the corresponding grading is called the pure ghost number $\#\pg$. This space is equipped with a natural differential $\gamma$, and the cohomology of the corresponding complex (the Chevalley-Eilenberg complex) is the space of invariant functions on $\E(M)$. 
 
The Batalin-Vilkovisky algebra $\BV(M)$ is now the alternating tensor algebra of graded derivations on the Chevalley-Eilenberg algebra. It has two gradings: ghost number $\#\gh$ and antifield number $\#\af$. Functionals of physical fields have both numbers equal to 0. Functionals of ghosts have a $\#\gh=\#\pg$ and  $\#\af=0$. All the derivations have a non-zero antifield number and $\#\gh=-\#\af$.  The space $\BV(M)$ is equipped with a graded generalization of the Schouten bracket.
One can extend this bracket to the level of natural transformations and obtain in this way an odd  graded Poisson bracket on the space of generalized Lagrangians. From the definition of the Koszul operator we know that it can be written as the antibracket with the original action $S$, i.e. 
\be\label{Koszul}
\delta_S F=\{F,L_M(f)\},\ F\in \BV(M),\,f\equiv 1\ \textrm{on }\supp\, F\,.
\ee
Note that since neither our spacetime nor the supports of field configurations are assumed to be compact, the differential $\delta_S$ is not inner with respect to the antibracket. Nevertheless one can use (\ref{Koszul}) to write it locally as $\{.,L_M(f)\}$ with a sufficiently chosen test function $f$. This issue is thoroughly discussed in \cite{FR}, where the significance of a category theoretical  formulation is stressed. To simplify the notation we write from now on $\delta_S F=\{F,S\}$ instead of  (\ref{Koszul}). In a similar manner one can find a natural transformation $\theta$, that implements the Chevalley-Eilenberg differential:  $\gamma F=\{.,\theta\}$. In the ``closed algebra'' situation the total BV operator is simply defined as the sum of these two differentials
 \[
 s\doteq\{.,S+\theta\}
 \]
 We have $s^2=\{\cdot,\{S+\theta,S+\theta\}=0$. The cohomology of $s$ is the space of invariant functionals on the space of solutions. One usually enlarges the complex without changing its cohomology (by adding auxiliary fields like antighosts, Nakanishi-Lautrup fields etc.) in such a way that one can find an element $\psi$ (the gauge fixing fermion) with the property that the field equation for the gauge fixed action $S+\{\psi,\theta\}$ has a well posed Cauchy problem.  All these auxiliary objects are incorporated into the structure together with their antifields (graded derivations). In \cite{FR} it was shown that the resulting space $\BV(M)$  can be treated as a space of smooth maps from the configuration space $\E(M)$ into a certain graded algebra. For these smooth maps we can again formulate regularity condition, analogous to those formulated for $\F(M)$. In particular we distinguish the spaces of local maps $\BV_\loc(M)$ and regular maps $\BV_\reg(M)$. In order to transport the structure of the BV complex into the quantum theory we first split $S=S_0+V$ into a quadratic functional $S_0$ with $\#\af=0$ and the rest. Again, to keep
focused on the essential structure, we want to discuss some preliminary steps already on the level of the nonrenormalized time-ordered product. 

Let us first consider $\BV_\reg(M)$. 
The algebra $\BV_\reg(M)$ contains also functionals of Fermionic fields (see \cite{Rej} for a detailed discussion of such objects), so some additional signs appear in the formulas used in the previous section. The operator $\Lap$ in the graded case is defined as:
\[
\Lap X=\sum\limits_\alpha(-1)^{|\ph^\al|(1+|X|)}\int  \frac{\delta^2 X}{\delta\ph_\alpha^\ddagger(x)\delta\ph^\alpha(x)}\qquad ,\  X\in\BV_\reg(M)\,,
\]
where $|.|$ denotes the ghost number $\#\gh$ and $\alpha$ runs through all the field configuration types of the theory, i.e. physical fields, ghosts, antighosts, etc. To simplify the notation we will denote the full multiplet just by $\ph$ and its components by $\ph^\al$.
 Time ordered products of graded functionals (i.e. elements with $\#\af=0$) can be defined with the use of formula (\ref{Tproduct}), were the functional derivatives in $\DD$ with respect to graded field configurations have to be understood as the left derivatives and corresponding sign rules appear. Time ordered products of derivations ($\#\af>0$)  are defined similar to the scalar case, i.e. by means of (\ref{timeordvf}).
 The antibrackets $\{.,.\}_{\TT}$ and  $\{.,.\}_{\star}$ are simply given by formulas (\ref{Delta:Tbracket}), (\ref{Delta:star:bracket}) with the graded version of $\Lap$ defined above.

With these structures at hand we want now to discuss the gauge fixing. 
 Our starting point is a classical Lagrangian, where a suitable canonical transformation was performed, so that the term of $\#\af=0$, quadratic in fields, induces a normally hyperbolic system of equations. This is the free part of the Lagrangian and we use it to define the free time-ordered product $\TT$. We denote the corresponding free action by $S_0$. The quantum Koszul operator associated with this action is $\{.,S_0\}_\star$. For details concerning the gauge fixing in the classical case we refer to \cite{FR,Barnich:1999cy}. Now we want to switch on the interaction. 

The interacting term of our action has to be chosen with some caution. We don't want to use any physical interaction yet, since these are local and therefore the nonlinear part would not be an element of $\BV_\reg(M)$. Instead we consider for the moment some other functional $V\in\TT(\BV_\reg(M))$ with ghost number $\#\gh=0$ which also contains antifields. In the spirit of perturbation theory we want to construct the interacting fields from the free ones using an appropriate intertwining map. A simple generalization from the scalar case suggests to use the map $R_V$, defined by (\ref{RV}). The quantum Koszul map has to be transformed as well. We define the \textit{\textbf{quantum BV operator}} $\hat{s}$ as the deformation of $\{.,S_0\}_\star$ under the action of $R_V$:
\be\label{intertwining:s}
\hat{s}\doteq R_V^{-1}\circ\{.,S_0\}_\star\circ R_V\,.
\ee
It is clear that $\hat{s}$ is a derivation with respect to the interacting star product $\star_V$. Moreover we can characterize the cohomology of $\hat{s}$ knowing the  one of $\{.,S_0\}_\star$. 

The natural question to ask now is, what will happen, if we change the gauge-fixing Fermion. In other words we want to perform again a canonical transformation $\alpha_\psi$ and see how $R_V(F)$ is changing. We choose the new gauge-fixing Fermion $\psi$ as an element of $\TT(\BV_\reg(M))$ with $\#\gh=-1$. Assume that $\psi$ doesn't contain antifields. Just like in the classical case \cite{FR}, first we define an automorphism of the algebra $\TT(\BV_\reg(M))$ by
\be\label{gfermionq}
\alpha_{\psi}(X):=\sum_{n=0}^{\infty}\frac{1}{n!}\underbrace{\{\psi,\dots,\{\psi}_n,X\}_{\TT}\dots\}_{\TT}=\TT(\alpha_{\TT^{\minus}\psi}(\TT^{\minus}X))\, .
\ee
In the second step of the gauge fixing procedure we set all the elements with $\#\ta>0$ in $\TT(\BV_\reg(M))$ to 0. Now we want to compare the original interacting field $R_V(F)$ with the one arising from the  $\alpha_{\lambda\psi}$-transformed free algebra, i.e. with  $R_{\alpha_{\la\psi}(V)}(\alpha_{\la\psi}(F))$.  Let us denote $\tilde{F}\doteq\alpha_{\la\psi}(F)$, $\tilde{V}\doteq\alpha_{\la\psi}(V)$. Similar to the standard approach to BV-quantization (see for example \cite{Henneaux:1992ig}) we want now to provide conditions, which assure that the S-matrix $e^{\tilde{V}}$ and the interacting field $R_{\tilde{V}}(\tilde{F})$ are independent of $\psi$, modulo terms that vanish on-shell.
 This can be formulated as:
\be\label{inS}
\frac{d}{d\lambda}\,\left(e_{\sst{\TT}}^{i\tilde{V}/\hbar}\right)\os 0\,,
\ee
and 
\be\label{indepobs}
\frac{d}{d\lambda}\,\left(e_{\sst{\TT}}^{i\tilde{V}/\hbar}\right)^{\star-1}\star \left(e_{\sst{\TT}}^{i\tilde{V}/\hbar}\T \tilde{F})\right)\os 0\,,
\ee
where ``o.s.'' means ``on shell'', i.e. modulo the ideal generated by the equations of motion derived from $S_0$. We start with the first of these formulas. Since $\frac{d}{d\lambda}\,e_{\sst{\TT}}^{i\tilde{V}/\hbar}=\left\{\psi,e_{\sst{\TT}}^{i\tilde{V}/\hbar}\right\}_{\TT}$, the condition (\ref{inS}) can be written as:
\be\label{indepS}
\left\{\psi,e_{\sst{\TT}}^{i\tilde{V}/\hbar}\right\}_{\TT}\os 0\,.
\ee
We can rewrite the left hand side of this equation using the identity (\ref{derivation:T}). We obtain the following condition:
\be\label{gaugeindep}
\{\psi\T e_{\sst{\TT}}^{i\tilde{V}/\hbar},S_0\}_\star-\{\psi,S_0\}_\star\T e_{\sst{\TT}}^{i\tilde{V}/\hbar}+\psi\T \{e_{\sst{\TT}}^{i\tilde{V}/\hbar},S_0\}_\star
\os0\,.
\ee
The second term vanishes, since both $S_0$ and $\psi$ don't depend on antifields. Note also that the first term is an element of the ideal of equations of motion. Therefore a sufficient condition to fulfill (\ref{indepS}) on-shell is
\be\label{suff:condition0}
\{e_{\sst{\TT}}^{i\tilde{V}/\hbar},S_0\}_\star=0\,.
\ee
This equation still depends on $\psi$, since $\tilde{V}=\alpha_{\lambda\psi}(V)$. To amend it we first apply
the equation (\ref{tkoszul2}) to obtain
\[
\{e_{\sst{\TT}}^{i\tilde{V}/\hbar},S_0\}_{\TT}-i\hbar\Lap(e_{\sst{\TT}}^{i\tilde{V}/\hbar})=0\,.
\]
From (\ref{Delta:bracket2}) it follows that $\Lap(\al_{\la\psi}X)=\al_{\la\psi}(\Lap X)$, so in particular $\Lap(e_{\sst{\TT}}^{i\tilde{V}/\hbar})=\al_{\la\psi}(\Lap (e_{\sst{\TT}}^{iV/\hbar}))$.  Using the Leibniz rule for $\{.,S_0\}_{\TT}$  and the fact that both $S_0$ and $\psi$ don't contain antifields we can rewrite (\ref{suff:condition0}) as:
\[
\alpha_{\la\psi}\left(\{e_{\sst{\TT}}^{i{V}/\hbar},S_0\}_{\TT}-i\hbar\Lap(e_{\sst{\TT}}^{i{V}/\hbar})\right)=\al_{\la\psi}(\{e_{\sst{\TT}}^{iV/\hbar},S_0\}_\star)=0\,.
\]
Therefore a sufficient condition we are looking for is:
\be\label{suff:condition}
\{e_{\sst{\TT}}^{iV/\hbar},S_0\}_\star=0\,.
\ee
This is the so called \textbf{\textit{quantum master equation}} (\qme). The above discussion also shows that if it is fulfilled for some particular choice of the gauge fixing Fermion, then it is fulfilled for all.
To write (\ref{suff:condition}) in a more commonly known form we use the fact that
the operator $\Lap$ acting on the exponential function produces: $\Lap(e_{\sst{\TT}}^{iV/\hbar})=\frac{i}{\hbar}(\Lap V+\frac{i}{2\hbar}\{V,V\})\T e_{\sst{\TT}}^{iV/\hbar}$. We arrive finally at the condition:
\[
\{ V,S_0\}_{\TT}+\frac{1}{2}\{V,V\}_{\TT}-i\hbar\Lap V=0\,.
\]
Using  the fact, that $S_0$ doesn't contain antifields, we can write the above result in the form of the {\qme} known from the literature.
\be\label{QME0}
\frac{1}{2}\{S_0+V,S_0+V\}_{\TT}=i\hbar\Lap (S_0+V)\,.
\ee
Note that this is exactly the same condition, which is used in the path integral formalism to assure the gauge independence of the gauge-fixed ``measure'' \cite{Henneaux:1992ig}. Using the {\qme} we can now write the BV operator defined in (\ref{intertwining:s}) in a more explicit form:
\be\label{QBV0}
\hat{s}X=e_{\sst{\TT}}^{-i V/\hbar}\T\left(\{e_{\sst{\TT}}^{ iV/\hbar}\T X,S_0\}_{\star}\right)\,.
\ee
In section \ref{renormBV} we will show that this expression for the quantum BV operator can be generalized to renormalized time-ordered products and no divergences appear. 

Using the same reasoning as for the {\qme}, by manipulating expression (\ref{indepobs}), we can conclude that if  (\ref{suff:condition}) holds, then the condition that $R_{\tilde{V}}(\tilde{F})$ on-shell is independent of the gauge fixing can be written as:
\be
\left(e_{\sst{\TT}}^{i\tilde{V}/\hbar}\right)^{\star-1}\star\left(\frac{d}{d\lambda}\,e_{\sst{\TT}}^{i\tilde{V}/\hbar}\T \tilde{F}\right)=\left(e_{\sst{\TT}}^{i\tilde{V}/\hbar}\right)^{\star-1}\star\left\{\psi,e_{\sst{\TT}}^{i\tilde{V}/\hbar}\T \tilde{F}\right\}_{\TT}\os 0\,.
\ee
Again we can rewrite it with the use of (\ref{derivation:T}) to obtain:
\[
\left(e_{\sst{\TT}}^{i\tilde{V}/\hbar}\right)^{\star-1}\star\left(\{\psi\T e_{\sst{\TT}}^{i\tilde{V}/\hbar}\T \tilde{F},S_0\}_\star+\psi\T \{e_{\sst{\TT}}^{i\tilde{V}/\hbar}\T \tilde{F},S_0\}_\star\right)\os 0\,.
\]
This is guaranteed if
\[
\left(e_{\sst{\TT}}^{i\tilde{V}/\hbar}\right)^{\star-1}\star\left(\psi\T \{e_{\sst{\TT}}^{i\tilde{V}/\hbar}\T \tilde{F},S_0\}_\star\right)=0\,.
\]
We can rewrite this condition using the definition of $R_{\tilde{V}}$ and of the quantum BV operator (\ref{intertwining:s}). We obtain
\[
R_{\tilde{V}}(\psi\T \hat{s}_{\tilde{V}}\tilde{F})=0\,,
\]
where by $\hat{s}_{\tilde{V}}$ we denoted the quantum BV defined by (\ref{intertwining:s}) with the interaction term $\tilde{V}$.
Using a similar reasoning as for the {\qme}, we can rewrite the above formula as:
\[
R_{\tilde{V}}(\psi\T \al_{\la\psi}(\hat{s}_{V}F))=0\,,
\] 
Therefore, if $F$ is in the cohomology of $\hat{s}_V$, then $ \hat{s}_VF=0$ and $R_{\tilde{V}}(\tilde{F})-R_V(F)\in\im(\{\cdot,S_0\}_\star)$.
Moreover it is clear that the cohomologies for the operators  $\hat{s}_V$ and  $\hat{s}_{\tilde{V}}$ are equivalent.
 
The quantum BV operator can be written in a more commonly used form with the use of equation (\ref{tkoszul2}) and the properties of $\Lap$:
\[
\hat{s}F=\{F,S_0+V\}_{\TT}-i\hbar \Lap F\,.
\]

To close this section we want to reflect a while on the question, whether one can add to the free Lagrangian a term that  contains antifields. Let us denote it by $\theta_0\in\TT(\BV_\reg(M))$. Of course it has to be linear both in fields and antifields. The full extended action takes the form $S_0+\theta_0+\lambda(\theta_1+S_1)=S_0+\theta_0+V$, where $\theta_1$ is linear in antifields, and $S_1\in\TT(\BV_\reg(M))$. We can interpret $\theta_0$ as the free BRST operator. The 0-th order in the coupling constant of equation (\ref{QME0}) is a statement that
\be\label{Stheta}
 \{S_0,\theta_0\}_{\TT}=0\,.
\ee
From this property we obtain:
\be\label{theta0}
\left\{e_{\sst{\TT}}^{i V/\hbar}\T X,\theta_0\right\}_\TT=\left\{e_{\sst{\TT}}^{iV/\hbar}\T X,\theta_0\right\}_\star\,.
\ee
We provide the proof of above relation in the appendix \ref{theta:proof}.
Using  (\ref{theta0}) we obtain a following formula:
\be\label{braketfull}
\{e_{\sst{\TT}}^{iV/\hbar}\T X,S_0+\theta_0\}_{\TT}=\{e_{\sst{\TT}}^{i V/\hbar}\T X,S_0+\theta_0\}_\star+i\hbar\Lap\left(e_{\sst{\TT}}^{i V/\hbar}\T X\right)\,.
\ee
In particular for $X=1$ we have:
\[
e_{\sst{\TT}}^{iV/\hbar}\T\left(\{V,\theta_0+S_0\}_{\TT}+\frac{1}{2}\{V,V\}_{\TT}-i\hbar\Lap (V)\right)=\{e_{\sst{\TT}}^{i V/\hbar},S_0+\theta_0\}_\star\,.
\]
The {\qme} for the free action (0-th order in $\lambda$) states that $\frac{1}{2}\{S_0+\theta_0,S_0+\theta_0\}_{\TT}=i\hbar\Lap( S_0+\theta_0)$, so the {\qme} for the full action $S_0+\theta_0+V$ guarantees that $\{V,\theta_0+S_0\}_{\TT}+\frac{1}{2}\{V,V\}_{\TT}-i\hbar\Lap (V)=0$ and we obtain:
\[
\{e_{\sst{\TT}}^{iV/\hbar},S_0+\theta_0\}_\star=0\,.
\]
Therefore the quantum BV operator can be alternatively written as:
\be\label{QBV2}
\hat{s}X=e_{\sst{\TT}}^{-iV/\hbar}\T\left(\{e_{\sst{\TT}}^{i V/\hbar}\T X,S_0+\theta_0\}_{\star}\right)\,.
\ee
We also obtain  the formulation of the on-shell gauge invariance of the S-matrix, which is closer to the one given in \cite{H}:
\[
\{e_{\sst{\TT}}^{iV/\hbar},\theta_0\}_\star=0\qquad\textrm{on shell}\,.
\]
To summarize, we have shown in this section, that important notions of the BV quantization have a natural interpretation in the language of {\paqft}. The problem we have to face right now is the generalization of these structures to more singular objects. As we already pointed out, the operator $\Lap$, which plays an important role in the BV-quantization is not well defined on local vector fields. This pathology results from the fact, that we were using the non-renormalized time-ordered product $\T$. Now we want to amend it, by means of renormalization. In the Section 4 we will show, that one can completely avoid any divergences, if one works with the renormalized time-ordered product from the very beginning,
provided the time-ordered product is equivalent to the pointwise product of classical physics. In the next section we will show that indeed the renormalized time ordered product of causal perturbation theory can be extended to a binary product with the desired properties.
\section{Renormalized time-ordered product}
\subsection{Causal perturbation theory}
In the previous section we considered only very regular objects which allowed us to present the general structure of the quantum theory without going into technical details. 
The relevant interactions, however, are local functionals. The crucial fact is now that after a properly defined normal ordering the operator product of local functionals is well defined. For Minkowski space this is the classical result of 
G{\aa}rding and Wightman \cite{WG64}; for generic globally hyperbolic spacetimes it was proven in \cite{BFK95}.

On Minkowski space, the normal ordering can be performed by the transformation $\al_{\Delta_1}\doteq\exp({\hbar\Ga_{\Delta_1}}):\F_\reg(\M)[[\hbar]]\rightarrow \F_\reg(\M)[[\hbar]]$ where $\Delta_1$ is the symmetric part of the Wightman 2-point function $\Delta_+=\frac{i}{2}\Delta+\Delta_1$. On a generic spacetime one uses instead of $\Delta_1$ an arbitrary Hadamard function $H$, i.e. a symmetric distribution in 2 variables such that the wave front set of $\frac{i}{2}\Delta+H$ satisfies the microlocal spectrum condition \cite{Rad}. This amounts to replacing the $\star$-product by an equivalent one
\be\label{star:H}
F\star_{\sst{H}} G\doteq\al_{H}(\al_{H}^{-1}(F)\star \al_{H}^{-1}(G)) \ .
\ee
The transformed operator product $\star_{\sst{H}}$ can now be extended to the space of microcausal functionals  $\F_\mc(M)$. Here a functional is called microcausal if it is smooth and if the wave front set of its functional derivatives does not contain elements $(x_1,\dots,x_n;k_1,\dots,k_n), k_i\in T_{x_i}^*M, i=1,\dots,n$ where all $k_i$ are in the closed forward lightcone or all in the closed backward lightcone. In particular, all $\star_{\sst{H}}$-products of local functionals are microcausal. See \cite{BF0} for more details.

In the same way we can also transform the time ordered product by replacing the time ordering operator $\TT$ by $\alpha_H\circ\TT$. On Minkowski space one may choose $H=\Delta_1$; this amounts to replacing the Dirac propagator $\Delta_D$ in the definition of $\TT$ by the Feynman propagator $\Delta_F$. 

The arising new time ordered product is still not well defined on local functionals due to the larger wave front sets of $i\Delta_D+H$ at coinciding points. This phenomenon is nothing else than a position space version of the well known ultraviolet divergences of perturbative quantum field theory.
The strategy of causal perturbation theory to deal with this problem is the following one: One uses the fact that the time ordered product coincides with the operator product for time ordered support of the factors. One then constructs a family of multilinear symmetric maps $\TT_n:\F_\loc(M)^n\to\F_\mc(M)[[\hbar]]=:\fA_\mc(M)$ with
\be
\supp\TT_n(F_1,\dots,F_n)\subset\bigcup\supp F_i
\ee
and the causal factorization rule \eqref{causalfactorization}. The map $\TT_1$ can be chosen as the identity, but on curved spacetimes this choice has bad covariance properties, so one better chooses
$\TT_1=e^{\Gamma_w}$ where $w$ is the smooth part in the Hadamard function,
\be
H=\frac{u}{\sigma}+v\ln\sigma+w
\ee 
with $\sigma(x,y)$ denoting the square of the length of the geodesic connecting $x$ and $y$ and with geometrical determined smooth functions $u$ and $v$. $\TT_1$ is up to the choice of a length scale uniquely fixed by this formula. See \cite{HW} for more details.

The maps $\TT_n$ can now inductively be constructed, and $\TT_n$ is uniquely fixed by the lower order maps $\TT_k$, $k<n$, up to the addition of an $n$-linear  map
\be
Z_n:\F_\loc(M)^n\to\F_\loc(M)[[\hbar]]=:\fA_\loc(M) \ ,
\ee
which describes the possible finite renormalizations.
We may now define time ordered products of $n$ elements of $\fA_\loc(M)$ by
\be\label{iterated-time-ordered-product}
A_1,\dots, A_n\mapsto\TT_n(\TT_1^{-1}A_1,\dots,\TT_1^{-1}A_n)\,.
\ee
\subsection{Associativity of the renormalized time-ordered \-product}\label{renprod}
In this subsection we show that the multilinear maps \eqref{iterated-time-ordered-product} arise from an iterated binary associative product $\TR$. 
The crucial observation is that an $n$-tuple of local functionals $F_1,\dots,F_n$ which vanish on some field configuration (say $\ph=0$ in case the configuration space is a vector space) is uniquely encoded in its pointwise product $F_1\cdots F_n$. The pointwise product is termed an $n$-local functional, and we may consider 
$\TT_n$ as a map on $n$-local functionals. In the following we restrict ourselves to the case where the configuration space is a vector space. 

Let $\F^{(0)}_\loc(M)$ be the space of local functionals which vanish at $\ph=0$, and let  $S^\bullet\F^{(0)}_\loc(M)$ denote the space of symmetric tensor powers of $\F^{(0)}_\loc(M)$. The pointwise multiplication $m$ maps $S^\bullet\F^{(0)}_\loc(M)$ onto the space of multilocal functionals $\F(M)$.
\begin{theorem}
The multiplication $m:S^\bullet\F^{(0)}_\loc(M)\to\F(M)$ is bijective.
\end{theorem}
\begin{proof}
By definition of the space of multilocal functionals $\F(M)$ the map $m$ is surjective.
To prove that it is also injective, let $F=\bigoplus\limits_{k=0}^nF_k \in S^\bullet\F_\loc^{(0)}(M)$, $F_k\in S^k\F_\loc^{(0)}(M)$, with $m(F)=0$. It follows that also the $n$-fold derivative of $m(F)$ is equal to $0$. Let us take $x_1,\ldots, x_n\in M$ such that $x_i\neq x_j$ for $i\neq j$. Then only $F_n$ contributes to the derivative due to the support property of $k$-local functionals with $k<n$, and we obtain
\[
\frac{\delta^n(m(F))}{\delta\ph(x_1)\ldots\delta\ph(x_n)}(\ph)=n!\frac{\delta^n F_n(\ph_1=\ph,\ldots,\ph_n=\ph)}{\delta\ph_1(x_1)\ldots\delta\ph_n(x_n)}=0\,.
\]
We know that 
\be
\frac{\delta^n F_n(\ph_1,\ldots,\ph_n)}{\delta\ph_1(x_1)\ldots\delta\ph_n(x_n)}
\ee
is a function of $x_1,\ldots,x_n$ which depends on the field configurations $\ph_1,\dots,\ph_n$ only via their jets $j_{x_i}(\ph_i)$ at the points $x_i$, $i=1,\dots,n$.
Let us now take arbitrary $\ph_1,\ldots,\ph_n$ and define a smooth partition of unity $1=\sum\limits_{i=1}^n\chi_i$, where  $\supp\,\chi_i\cap\{x_j;j\neq i\}=\varnothing$. Now we set $\ph=\sum\limits_{i=1}^n\chi_i\ph_i$, thus $j_{x_i}(\ph)=j_{x_i}(\ph_i), i=1,\dots,n$ and it follows that:
\be\label{multilical:0}
\frac{\delta^n F_n(\ph_1,\ldots,\ph_n)}{\delta\ph_1(x_1)\ldots\delta\ph_n(x_n)}=0\,.
\ee
Up to now, we know that this holds for pairwise distinct points $x_i$. But by the definition of $S^n\F_\loc^{(0)}(M)$, the above derivative is an everywhere defined smooth function of $x_1$, \ldots, $x_n$, so it is equal to $0$ also for coinciding points.

Again by the definition of $S^n\F_\loc^{(0)}(M)$, the functional $F_n$ vanishes if one of its arguments is the distinguished configuration $\ph=0$. Hence
\begin{align}
F_n(\ph_1,\dots,\ph_n)&=\sum_{I\subset\{1,\dots,n\}}(-1)^{|I|}F_n(\psi_1,\dots,\psi_n)|_{\psi_i=0\text{ for }i\in I,\psi_i=\ph_i\text{ for }i\not\in I} \\
&=\int_{[0,1]^n}d\lambda_1\dots d\lambda_n\left\langle \frac{\delta^n F_n(\lambda_1\ph_1,\dots,\lambda_n\ph_n)}{\delta\ph_1\dots\delta\ph_n},\ph_1\otimes\dots\otimes\ph_n\right\rangle\,.
\end{align}
Thus $F_n=0$. Iterating the argument gives $F_k=0$ for $1\le k\le n$. Hence $F$ is a constant and $m(F)(\ph)=F$. This implies $F=0$.
\end{proof} 
Let $\beta=m^{-1}$. We now define the renormalized time ordering operator on the space of multilocal functionals $\F(M)$ by
\be
\TTR:=(\bigoplus_n  \TT_n)\circ\beta \ .
\ee  
This operator is a formal power series in $\hbar$ starting with the identity, hence it is injective.
The renormalized time ordered product is now defined on the image of $\TTR$ by
\be
A\TRH B\doteq\TTR(\TTR^{\minus}A\cdot\TTR^{\minus}B)\,.
\ee
This product is equivalent to the pointwise product and is in particular associative and commutative. 
Note that $\TRH$ is well defined not on the full $\fA(M)$, but on a smaller space (which is invariant under the renormalization group action) namely $\Dcal_{\TTR}(M)\doteq\TTR(\F(M))[[\hbar]]$. Moreover, the $n$-fold $\TR$-product of local functionals coincides with the $n$-linear map of causal perturbation theory (equation \eqref{iterated-time-ordered-product}), namely one has:
\[
A_1\TR\dots\TR A_n=\TT_n(\TT_1^{-1}A_1,\dots,\TT_1^{-1}A_n)\,,
\]
when $A_1,\dots, A_n\in \fA_{\loc}(M)$.

Similarly as in section \ref{general} we can use the renormalized time ordering operator $\TTR$ to bring classical structures to the quantum world. 
In particular we can define  the time ordering of multilocal vector fields. Let $X\in\V(M)$, then we define
\be\label{rprodv}
\TTR X\doteq \int dx \TTR( X(x))\frac{\delta}{\delta\ph(x)}\,.
\ee
Since $\TR$ is now defined as a full product on $\Dcal_{\TTR}(M)$, we can repeat the reasoning from section \ref{general} and define the  $\TTR$-transformed Koszul operator with the renormalized time-ordered product in place of $\T$. Let $S\in\F_\loc(M)$ be the free action functional. The renormalized time ordered Koszul map is defined as
\[
\delTR\doteq\TTR\circ\delta_{\mathcal{T}^{\minus}S}\circ\TTR^{-1}\ .
\]
Clearly it is a well defined object and no divergences are present. We can also define the time-ordered antibracket:
\[
\{X,Y\}_{\sst{\TTR}}=\TTR\{\TTR^{-1}X,\TTR^{-1}Y\}\ .
\]

The definitions introduced above allow us to provide a mathematically rigorous interpretation of the renormalized quantum BV operator and the renormalized {\qme}. Before we turn to this task we want to make some remarks about the problems encountered in other approaches to the BV quantization. Note that the source of divergences in expression (\ref{QMO}) is the operator $\Lap$, which is ill defined on local vector fields.  In the standard approach this is solved by using an appropriate regularization scheme. Instead we show, that this problem can be completely avoided if  we work with renormalized time ordered products $\TTR$ from the very beginning. We shall follow now all the steps outlined in \ref{general} and see what is changing when we take the renormalization into account.
\section{BV formalism and renormalization}\label{renormBV}
\subsection{The renormalized quantum BV operator and the quantum master equation}
Now we have all the tools needed to introduce the interacting \textbf\textit{renormalized quantum BV operator}. We start with the classical algebra $\BV(M)$ underlying the BV-complex. It consists of functionals (elements with $\#\ta=0$) and derivations ($\#\ta>0$). The main difference with respect to the scalar case is the appearance of a grading. This implies that axioms for the time ordered products have to be modified by introducing appropriate sign rules. For example time ordered products of ghosts are antisymmetric instead of symmetric. Time ordered products of derivations are defined in the same way as time-ordered products of vector fields, i.e. by means of (\ref{rprodv}). The resulting quantum algebra of free fields will be again denoted by $\fA(M)$ and $\TTR(\BV(M))$ is its subset.

With these considerations in mind we can now set to define the renormalized BV operator. We can do it in a similar way to the non-renormalized case by using the expression (\ref{QBV0}) or  (\ref{QBV2}) with $\TT$ replaced by $\TTR$, namely:
\be\label{QBVr}
\hat{s}(X)=e_{\sst{\TTR}}^{-iV/\hbar}\TR\{e_{\sst{\TTR}}^{iV/\hbar}\TR X,S_0\}_{\star}\,,
\ee
where $V, X\in\TTR(\BV_\loc(M))$. Note that $\{.,.\}_\star$ is not defined on the full space of microcausal functionals, because of the singularities of $\Delta_+$. We know however that it is well defined, if one of the arguments is regular or equal to $S_0$. The renormalized time-ordered antibracket $\{.,.\}_{\TTR}$ on the other hand makes sense for all multilocal functionals.

To understand better the expression for $\hat{s}$, we shall use the anomalous Master Ward Identity ({\mwi}) \cite{BreDue,DueBoas}. In our formalism it takes the following form:
\begin{proposition}
Let $V\in\TTR(\BV_\loc(M)$. Then there exists (in the sense of a formal power series in $V$) a map 
\be
\Lap_r:\TTR(\BV_\loc(M))\to \fA_\loc(M)
\ee
with $\supp(\Lap_r(V))\subset \supp V$ such that
\be\label{MWIV}
\int \Big(e_{\sst{\TTR}}^{iV/\hbar}\TR \frac{\delta V}{\delta\ph^\ddagger(x)}\Big)\star\frac{\delta S_0}{\delta\ph(x)}=e_{\sst{\TTR}}^{iV/\hbar}\TR\big(\tfrac{1}{2}\{V+S_0,V+S_0\}_{\TTR}-i\hbar\Lap_r(V)\big)\,,
\ee
\end{proposition}
The proof can be found in \cite{BreDue,DueBoas,H}. Using $\Lap_r$ one can now define a linear operator $\Lap_V:\TTR(\BV_\loc(M))\to \fA_\loc(M)$ by means of
\[
\Lap_V(X)\doteq \frac{d}{d\lambda}\Big|_{\lambda=0}\Lap_r(V+\lambda X),\quad X\in\TTR(\BV_\loc(M)\,.
\]
It follows that
\be\label{MWI}
\int (e_{\sst{\TTR}}^{iV/\hbar}\TR \frac{\delta X}{\delta\ph^\ddagger(x)})\star\frac{\delta S_0}{\delta\ph(x)}=e_{\sst{\TTR}}^{iV/\hbar}\TR(\{X,V+S_0\}_{\TTR}-i\hbar\Lap_V(X))\,,
\ee
and $\supp\Lap_V(X)\subset \supp X\cap \supp V$. If $X$ is of the form $X=\int X(x)\tfrac{\delta}{\delta \ph(x)}$, and both $X(x)$ and $V$ don't contain antifields, formula \eqref{MWI} reduces to the case studied in  \cite{BreDue,DueBoas} and $\Lap_V(X)$ is the so called ``anomaly''\footnote{In the original paper \cite{BreDue} the anomaly term is denoted by $\Lap_X(V)$. We use an opposite convention since it resembles more the notation used for the Laplacian operator $\Lap$ defined on the regular vector fields.}.

Comparing formula \eqref{MWIV} with (\ref{QME0}) we see that the \textsc{mwi} provides us with means to formulate the \textit{\textbf{renormalized quantum master equation}}. The singular operator $\Lap$, which was independent of the interaction is now replaced by the finite expression, which in turn depends on $V$ in a nonlinear way.
We obtain the renormalized {\qme} in the form:
\be\label{QMEr}
\frac{1}{2}\{V+S_0,V+S_0\}_{\TTR}=i\hbar\Lap_r(V)\,.
\ee 
Just like in the non-renormalized case, fulfilling the {\qme} (\ref{QMEr}) is equivalent to the invariance of  the extended S-matrix under the quantum Koszul operator. This guarantees that the equation (\ref{intertwining:s}) is fulfilled also in the renormalized case, i.e.:
\be\label{intertwining:s:r}
\{.,S_0\}_\star\circ R_V=R_V\circ\hat{s}\,,
\ee
where  $R_V(G)=(e_{\sst{\TTR}}^{iV/\hbar})^{-1,\star}\star (e_{\sst{\TTR}}^{iV/\hbar}\TR G)$. We see that the interpretation of $\hat{s}$  as the $R_V$-transformed $\{.,S_0\}_\star$ carries over also to the renormalized theory. 

Using (\ref{QMEr}) and \eqref{MWI}, we obtain, for an arbitrary element $\TTR X\in\TTR(\BV(M))$, a following simple expression for the renormalized quantum BV-operator:
\[
\hat{s}X=\{X,V+S_0\}_{\TTR}-i\hbar\Lap_V(X)\,.
\]
We will now take a closer look at the nature of the anomaly term and try to understand it better, by formulating certain consistency conditions. First we note that in equation (\ref{QBVr}) the star product with $\frac{\delta S_0}{\delta\ph(x)}$ or $\frac{\delta S_0}{\delta\ph^\ddagger(x)}$ can be also replaced by the pointwise product and therefore:
\[
\{\{e_{\sst{\TTR}}^{iV/\hbar}\TR X,S_0\}_{\star},S_0\}_\star=0
\]
This implies that:
\[
\hat{s}^2( X)=0
\]
From this condition and the classical master equation for $S_0+V$ it follows that
\[
\{\Lap_V(X),V+S_0\}_{\TTR}+\Lap_V(\{X,V+S_0\}_{\TTR})-i\hbar\Lap_V(\Lap_V(X))=0
\]
Note that $\{.,V+S_0\}_{\TTR}$ is just the classical BV operator $s=\{.,\TTR^{\minus}(V+S_0)\}$ transported to the quantum algebra by means of $\TTR$. Therefore, if $\TTR^{\minus}X$ is invariant under $s$, then also $\{X,V+S_0\}_{\TTR}=0$ and we obtain a condition analogous to the Wess-Zumino consistency condition \cite{Wess}:
\be\label{constcond2}
\{\Lap_V(X),V+S_0\}_{\TTR}=i\hbar\Lap_V(\Lap_V(X))
\ee
Similarly, applying $\{.,S_0\}_\star$ twice on $e_{\sst{\TTR}}^{iV/\hbar}$ itself and using the nilpotency of $s$ we obtain:
\[
\{\Lap_r(V),V+S_0\}_{\TTR}+\Lap_V(\{V+S_0,V+S_0\}_{\TTR})=i\hbar\Lap_V(\Lap_r(V))\,.
\]
If the classical master equation holds for the action $\TTR^{\minus}V$, it follows that
\be\label{constcond3}
\{\Lap_r(V),V+S_0\}_{\TTR}=i\hbar\Lap_V(\Lap_r(V))\,.
\ee
\subsection{Algebraic adiabatic limit}\label{adiablim}
In causal perturbation theory we work with interactions that are localized, but usually we have to deal with
interacting theories, where there is no natural cutoff. To circumvent this problem we can introduce it by replacing the coupling constant with a compactly supported function $f$. This cutoff can be then removed, using the construction called the algebraic adiabatic limit \cite{BDF}. It is done in the framework of locally covariant quantum field theory \cite{BFV}. In this section we review briefly the most important definitions used in this formalism.

Let $\fA(M)$ be the functor into the category of involutive algebras  that associates to $M$ the quantum  algebra $\fA(M)$. It has a subfunctor $\fA_\loc$ that associates to a spacetime the space of local observables $\fA_\loc(M)$. The generalized Lagrangian $L$ is defined in analogy to the definition in section \ref{scalar}, but we use the functor $\fA_\loc$ instead of $\F_\loc$.

More generally, following \cite{FR}, we can also include in our discussions the generalized Lagrangians of higher order. Let $\Nat( \D,\BV_\loc )$ denote the set of natural transformations and $\D^k$ is a functor from the category $\Loc$ to the product category ${\Vect}^k$, that assigns to a manifold $M$ a $k$-fold product of the test section spaces $\D(M)\times\ldots\times \D(M)$. Let $\Nat(\D^k,\fA_\loc)$ denote the set of natural transformations from $\D^k$ to $\fA_\loc$. We define extended Lagrangians $L\in Lgr$ to be elements of the space $\bigoplus_{k=0}^\infty \Nat(\D^k,\fA_\loc)$ satisfying: $\supp(L_M(f_1,...,f_n))\subseteq \supp f_1\cup...\cup\supp f_n$ and the additivity rule in each argument. We can introduce on $Lgr$ an equivalence relation similar to (\ref{equ}). We say that $L_1\sim L_2$, $L_1,L_2\in\Nat(\D^k,\fA_\loc)$ if:
\be\label{equ2}
\supp((L_1-L_2)_M(f_1,...,f_k))\subset \supp(df_1)\cup...\cup\supp(df_k),\ \forall f_1,...,f_k\in\D^k(M)
\ee
\subsection{{\qme} in the algebraic adiabatic limit and the renormalization group}\label{adiablimqme}
The idea to generalize the renormalization group to the level of natural transformations may seem to be a little bit abstract at the beginning. It is however very useful, if we want to have control on the cutoff needed to localize the interaction. In this section we will show that also the quantum master equation appears naturally in this setting. The idea is similar to the case of the classical master equation discussed in \cite{FR}. Working on the level of natural transformations we avoid problems with boundary terms arising from the cutoff function. 

We start with the classical master equation ({\cme}). In \cite{FR} it was discussed for natural transformations between the functors $\D$ and $\BV$. Let $L_0$ be the free generalized Lagrangian and $L_1$ the interaction term. Both are now to be understood as natural transformations between $\D$ and $\BV$. The classical master equation is formulated as the condition that:
\be\label{CME}
\{L_0+L_1,L_0+L_1\}\sim 0\,,
\ee 
with the equivalence relation defined in (\ref{equ2}). It guarantees the nilpotency of the BV operator $s$ which is defined by $sF=\{F,L_0+L_1(f)\}$, where $f\equiv1$ on $\supp F$, $F\in \BV(M)$. 

Assume that we are given a solution of the {\cme}. Now we want to transport this structure to the quantum algebra. We use the fact that from the construction performed in \cite{HW} follows that $\TT_1$ is a functor from $\F_\loc$ to $\fA_\loc$. This allows us to construct natural transformations $\TT_1({L_0})$ and   $\TT_1({L_1})$  from $\D$ to $\fA$. We denote the corresponding equivalence classes by $S_0$ and $S_1$ and it holds:
\be\label{CME0}
\{S_0+S_1,S_0+S_1\}_{\TTR}\sim 0\,.
\ee
This is the {\cme} transported to the quantized algebra. The quantum BV operator is defined as
\be\label{eQBVr}
\hat{s}(X)=e_{\sst{\TTR}}^{-i{L_1}_M(f_1)/\hbar}\TR\left(\{e_{\sst{\TTR}}^{i{L_1}_M(f_1)/\hbar}\TR  X,{L_0}_M(f)\}_{\star}\right)\,,
\ee
where $\supp\, X\subset\Ocal$ and $f,f_1\equiv 1$ on $\Ocal$. The quantum master equation is a statement that the S-matrix in the algebraic adiabatic limit is invariant under the quantum BV operator, i.e.:
\[
\supp\left(e_{\sst{\TTR}}^{-i{L_1}_M(f_1)/\hbar}\TR\left(\{e_{\sst{\TTR}}^{i{L_1}_M(f_1)/\hbar},{L_0}_M(f)\}_{\star}\right)\right)\subset \supp\, df\cup \supp\, df_1\,.
\]
Using  the {\mwi} we can see that this expression is again an element of $\fA_\loc(M)$, so the condition above can be also formulated on the level of natural transformations:
\be\label{eQMEr0}
e_{\sst{\TTR}}^{-iS_1/\hbar}\TR\left(\{e_{\sst{\TTR}}^{i S_1/\hbar},S_0\}_{\star}\right)\sim 0\,,
\ee
This is the extended quantum master equation. We can write it in a more explicit form using (\ref{MWI}). Note that the anomaly term $\Lap_{{L_1}_M(f)}({L_1}_M(f))$ is a natural transformation as well, so (\ref{eQMEr0}) is equivalent to:
\be\label{eQMEr1}
\frac{1}{2}\{S_0+S_1,S_0+S_1\}_{\TTR}-\Lap_{S_1}(S_1)\sim 0\,.
\ee
Note the resemblance of this condition to the classical master equation {\cme} (\ref{CME0}). The quantum BV operator can be now written as
\[
\hat{s}X=\{X,{L_0}_M(f)+{L_1}_M(f)\}_{\TTR}-\Lap_{{L_1}_M(f)}(X)\,,
\]
where $f\equiv 1$ on the support of $X\in\TTR(\BV(M))$. 

Now we want to see how the {\qme} and the quantum BV operator are transforming under the renormalization group.  

Let us recall the main theorem of renormalization, in the general form as proved in \cite{DF04,BDF} but adapted to our present formalism. It states that 2 different S-matrices $\Scal$ and $\hat{\Scal}$ are related by the formula
\be
\hat{\Scal}=\Scal\circ Z 
\ee
where $Z$ is an element of the St\"uckelberg-Petermann Renormalization Group $\Rcal$, i.e. it is a map from $\fA_\loc(M)$ to $\fA_\loc(M)$ which satisfies the conditions
\begin{enumerate}[{\bf Z 1.}]
\item $Z(0)  =0$,\label{Z0}
\item $Z^{(1)}(0)  =\id$,
\item $Z(A+B+ C)= Z(A+B)-Z(B)+Z(B+C)$ if $\supp(A)\cap\supp(C)=\emptyset$\label{Zloc1},
\item $\de Z/\de\varphi = 0$\label{Zindep}.
\end{enumerate}
If $\Scal$ is replaced by $\hat{\Scal}=\Scal\circ Z$ with a renormalization group element $Z$, then for an observable $F\in\fA(M)$ we obtain
\[
\hat{\Scal}_V(F)=\Scal_{Z(V)}(Z_V(F))
\]
where $Z_V(F)=Z(V+F)-Z(V)$ and it holds:
\begin{equation} 
\supp\, Z_V(F)\subset\supp\, F \ .\label{suppZ} 
\end{equation}

Now we show how the {\qme} is transforming under the action of $\Rcal$.
\begin{proposition}\label{renorm:qme}
Let $L_1$ be a natural Lagrangian that solves the {\qme} (\ref{eQMEr0}) for the renormalized time-ordered product $\TTR$. Let $Z\in\Rcal$ be the element of the renormalization group, which transforms between the $S$-matrices corresponding to $\TTR$  and ${\TTR}'$, i.e. $e_{\sst{\TTR}}^{{L_1}_M(f)}=e_{\sst{{\TTR}'}}^{Z({L_1}_M(f))}$. Then $Z(F)$ solves the {\qme} corresponding to ${\TTR}'$.
\end{proposition}
\begin{proof}
From the equation  (\ref{eQMEr0}) it follows that there exists a local element $A(f,f_1)\in\fA_\loc(M)$, depending on test functions $f$, $f_1$, such that $\supp A(f,f_1)\subset \supp\, df\cup \supp\,df_1$, such that:
\[
\{e_{\sst{\TTR}}^{i {L_1}_M(f_1)/\hbar},{L_0}_M(f)\}_{\star}=e_{\sst{\TTR}}^{-i{L_1}_M(f_1)/\hbar}\TR A(f,f_1)
\]
We can now transform both sides with the renormalization group element $Z$ to obtain:
\[
\{e_{\sst{{\TTR}'}}^{i Z({L_1}_M(f_1))/\hbar},{L_0}_M(f)\}_{\star}=e_{\sst{{\TTR}'}}^{-iZ({L_1}_M(f_1))/\hbar}\cdot_{{\TTR}'} \langle Z^{(1)}({L_1}_M(f_1)),A(f,f_1)\rangle\,.
\]
Using the property (\ref{suppZ}) of the renormalization group, we can conclude that 
\[
\supp \langle Z^{(1)}({L_1}_M(f_1)),A(f,f_1)\rangle\subset\supp A(f,f_1)\,.
\]
Hence:
\[
\supp\left(e_{\sst{{\TTR}'}}^{-iZ({L_1}_M(f_1))/\hbar}\cdot_{{\TTR}'}\{e_{\sst{{\TTR}'}}^{i Z({L_1}_M(f_1))/\hbar},{L_0}_M(f)\}_{\star}\right)\subset \supp\, df\cup \supp\, df_1\,.
\]
\end{proof}
We can see from the above proposition that the {\qme} is indeed a universal notion and transforms correctly under the renormalization group. A similar property can be shown for the $BV$ operator. To distinguish between operators corresponding to different interaction terms we denote by $\hat{s}_{\sst{S_1}}$ the quantum $BV$ operator defined for the action $S_1$ with respect to the time-ordering operator $\TTR$. 
For a different time ordering $\TTR'$ we obtain a corresponding  operator $\hat{s}'_{\sst{S_1}}$ in the form:
\be\label{s:hat:strich}
\int  (e^{iS_1/\hbar}\cdot_{\TTR'} X(x))\star\frac{\delta S_0}{\delta\ph(x)}= e_{\sst{\TTR'}}^{iS_1/\hbar}\cdot_{\TTR'}\hat{s}'_{\sst{S_1}}(X)\,,
\ee
On the other hand we know from the main theorem of renormalization 
that there exists an element $Z\in\Rcal$ such that the left hand side of the above formula can be written as:
\begin{multline*}
\int (e_{\sst{\TTR'}}^{i S_1/\hbar}\cdot_{\TTR'}X(x))\star\frac{\delta S_0}{\delta\ph(x)}=\int\, e_{\sst{\TTR}}^{iZ(S_1)/\hbar}\TR \langle Z^{(1)}(S_1),X(x)\rangle\star\frac{\delta S_0}{\delta\ph(x)}=\\
=e_{\sst{\TTR}}^{iZ(S_1)/\hbar}\cdot_{\TTR}(\hat{s}_{\sst{Z(S_1)}}\langle Z^{(1)}(S_1),X\rangle)\,,
\end{multline*}
Similarly we can rewrite the right hand side of (\ref{s:hat:strich}) as
\[
e_{\sst{\TTR'}}^{iS_1/\hbar}\cdot_{\TTR'}\hat{s}'_{\sst{S_1}}(X)=e_{\sst{\TTR}}^{iZ(S_1)/\hbar}\cdot_{\TTR}\langle Z^{(1)}(S_1),\hat{s}'_{\sst{S_1}}(X)\rangle\,.
\]
By comparing the above formulas we obtain:
\[
\hat{s}_{\sst{Z(S_1)}}\langle Z^{(1)}(S_1),X\rangle=\langle Z^{(1)}(S_1),\hat{s}'_{\sst{S_1}}(X)\rangle\,.
\]
Since it holds for arbitrary $X$, we can write the above relation as:
\be
\hat{s}_{\sst{Z(S_1)}}\circ Z^{(1)}(S_1)=Z^{(1)}(S_1)\circ\hat{s}'_{\sst{S_1}}\,.
\ee
This means that also the quantum BV operator transforms under the renormalization group in the natural way.

To end this section we want to discuss the problem of finding a solution to the {\qme}.
We start with a classical action, which satisfies the {\cme}, i.e. $\{S_0+S_1,S_0+S_1\}_{\TTR}=0$. For our renormalized time-ordering operator $\TTR$ we calculate the corresponding anomaly term $\Lap_{r}(S_1)$. In general it doesn't vanish, so the {\qme} will not be fulfilled. There are basically two possibilites to proceed. We can either redefine $\TTR$ using the renormalization freedom, or try to absorb $\Lap_{r}(S_1)$ into the action, by adding terms of higher order in $\hbar$. 
 The second way is more in the spirit of the original formulation of the BV formalism \cite{Batalin:1981jr,Batalin:1983wj,Batalin:1983jr}, so we follow this path first.  The cohomological problem can be formulated in the following way: we look for natural transformations $W_n$ such that $W=\sum_n \hbar^n W_n$, $W_0=S_1$ and
\be\label{W}
\frac{1}{2}\{W+S_0,W+S_0\}_{\TTR}-\Lap_{r}(W)\sim0
 \ee
holds. Let us expand $\Lap_{r}(W)$ as a power series in $\hbar$: $\Lap_{r}(W)=\sum_{k=0}^\infty \tfrac{\hbar^k}{k!}\Lap_{k}(W)$. It follows that the lowest order term in $\Lap_{r}(W)$ is $\Lap_{0}(S_1)$. Therefore, in the first order in $\hbar$, we obtain a condition:
 \be\label{firstorder}
 \{W_1,S_0+S_1\}_{\TTR}-i\Lap_0(S_1)\sim 0\,.
 \ee
From the consistency condition (\ref{constcond3}) in the first order in $\hbar$, we know that 
\[
\{\Lap_{0}(S_1),S_0+S_1\}_{\TTR}\sim0\,.
\]
Therefore the solution $W_1$ to (\ref{firstorder}) is governed by the cohomology of $s$ on the space of actions. To understand better this cohomological problem, recall that the action is an equivalence class of Lagrangians and these are in turn characterized by maps from $\D(M)$ to $\fA_\loc(M)$. Therefore calculating the cohomology of $s$ on the space of actions effectively amounts to calculate the cohomology of $s$ modulo $d$ on the space of local forms (polynomials of fields and their derivatives). Results in this direction were obtained in \cite{BBH,BarHenn,HennBar}. If the cohomology of $s$ turns out to be trivial, the existence of $W_1$ is guaranteed and we can insert it back to the equation (\ref{W}), which is now satisfied in the first order in $\hbar$. Next, we calculate the higher order terms and proceed inductively. To perform the induction step we assume that the {\qme} is fulfilled up to the order $n$. The consistency condition in the n-th order reads:
\[
\sum\limits_{k=0}^n\left(n\atop k\right)\{\Lap(W)\big|_{\textrm{order}\ k},W_{n-k}\}+\Lap\left(\{W,W\}\right)\big|_{\textrm{order}\ n}-i\Lap_W(\Lap(W))\big|_{\textrm{order}\ n-1}=0\,.
\]
Since the {\qme} is fulfilled in lower orders, the last two terms cancel and it follows that
\be\label{const:cond:n}
\sum\limits_{k=0}^n\left(n\atop k\right)\{\Lap(W)\big|_{\textrm{order}\ k},W_{n-k}\}=0\,.
\ee
The {\qme} in order $n+1$ is a condition that
\be\label{qme_np1}
\frac{1}{2}\sum\limits_{k=1}^{n}\binom{n+1}{k}\{W_k,W_{n+1-k}\}+\{W_{n+1},S_0+S_1\}-i\Lap(W)\big|_{\textrm{order}\ n}=0\,.
\ee
Using the graded Jackobi identity for $\{.,.\}$ and the {\qme} in orders lower than $n+1$, we can conclude that
\[
s\left(\frac{1}{2}\sum\limits_{k=1}^{n}\binom{n+1}{k}\{W_k,W_{n+1-k}\}-i\Lap(W)\big|_{\textrm{order}\ n}\right)=
\sum\limits_{k=0}^n\left(n\atop k\right)\{\Lap(W)\big|_{\textrm{order}\ k},W_{n-k}\}\,,
\] 
and this expression vanishes due to the consistency conditions \eqref{const:cond:n}. If the first cohomology of $s$ on the space of actions is trivial, then we can find $W_{n+1}$ such that  \eqref{qme_np1} is fulfilled and this proves the induction step.

In this way we can reduce the construction of $W$ to a strictly cohomological problem. Finding a solution $W$ of the {\qme} provides us with a map $S_1\mapsto W$. From the properties of $W$ it follows that there exists an element $Z$ of the renormalization group such that $Z(S_1)=W$, so we can write the {\qme} in the form:
\[ 
e_{\sst{\TTR}}^{-iZ(S_1)/\hbar}\TR\{e_{\sst{\TTR}}^{i Z(S_1)/\hbar},S_0\}_{\star}\sim 0 \ .
\]
From the main theorem of renormalization theory and proposition \ref{renorm:qme} it follows that there exists a time ordering operator ${\TTR}'$ such that:
\[
e_{\sst{{\TTR}'}}^{-iS_1/\hbar}\cdot_{\sst{{\TTR}'}}\{e_{\sst{{\TTR}'}}^{i S_1/\hbar},S_0\}_{\star}\sim 0\ .
\]
In this way we showed that the violation of the {\qme} can be also absorbed into the redefinition of the time-ordered product. This approach agrees with the one taken in \cite{H}.
\subsection{Relation to the regularized {\qme}} 
The construction of the renormalized quantum BV operator and the {\qme} we propose is completely independent of any regularization scheme, but it is interesting to see how our approach relates to those involving an explicit regularization. In particular  we want to make contact with the works of K. Costello \cite{Costello,CostBV,Cost2}.  Following \cite{BDF} we define the regularized time-ordered product corresponding to the scale $\Lambda$ as $\Tcal_\Lambda\doteq \exp(i\hbar \HL)$, where
\[
\HL=\frac{1}{2}\int dxdy(h_\Lambda-H)(x,y)\frac{\delta^2}{\delta\ph(x)\delta\ph(y)}\,,
\]
 and $h_\Lambda\xrightarrow{\Lambda\rightarrow\infty}H_F:=H+i\Delta_D$ in the sense of H\"ormander. 
 This provides a regularization of the Feynman like propagator $H_F$. The regularized S-matrix is now defined as $\Scal_\Lambda\doteq \exp_{\Tcal_\Lambda}$ and the regularized time-ordered Koszul operator is given by $\deL\doteq\TTL\circ\delta_{{\TTL}^{\minus}S}\circ{\TTL}^{\minus}$.  The regularized quantum BV operator\index{BV!operator!quantum regularized} is defined by replacing $\TT$ with a regularized time-ordered product in (\ref{QBV0}), i.e
\begin{multline}\label{RegBV0}
\hat{s}_\La X=e_{\sst{\TTL}}^{-i V/\hbar}\TL\left(\frac{\delta}{\delta\ph^\ddagger(x)}( e_{\sst{\TTL}}^{i V/\hbar}\TL X)\star\frac{\delta S_0}{\delta\ph(x)}\right)=\\
=e_{\sst{\TTL}}^{-i V/\hbar}\TL\{e_{\sst{\TTL}}^{i V/\hbar}\TL X,S_0\}_{\star}\,.
\end{multline}
The regularized quantum master equation\index{master equation!quantum regularized} can be understood as the condition that the regularized S-matrix is invariant under the quantum Koszul operator, i.e.:
\[
\{e_{\sst{\TTL}}^{iV/\hbar},S_0\}_{\star}=0
\]
 Let $V=V_0+\int  V_1(z)\frac{\delta}{\delta \ph(z)}$, where $V_0$ doesn't depend on antifields. We can write the regularized {\qme} explicitly using the fact that:
 \begin{eqnarray*}
 \deL(\Scal_\Lambda(V))&=&\TTL\delta_{{\TTL}^{\minus}S_0}\left(e^{i {\TTL}^{\minus}V/\hbar}\right)=m\circ e^{i\hbar \HL'}\left(\int \frac{\delta S_0}{\delta\ph(x)}\otimes\left(V_1(x)\TL e^{i V/\hbar}_{\TL}\right)\right)=\\
 &=&\int \frac{\delta S_0}{\delta\ph(x)}\left({\TTL}V_1(x)\TL e^{V}_{\TL}\right)+\TTL\left(i\hbar\Lap_\Lambda V-\frac{1}{2}\{V,V\}_\Lambda \right)\TL e^{V}_{\TL}\,,
 \end{eqnarray*}
 where by $\Lap_\Lambda$ we denoted the differential operator:
 \[
\Lap_\Lambda \doteq\int \!\! \frac{\delta^2 S_0}{\delta\ph(z)\ph(x)}(h_\Lambda-H)(x,y)\frac{\delta^2}{\delta\ph^\ddagger(z)\delta \ph(y)}\,,
\]
and $\{.,.\}_\Lambda$ is the scale $\Lambda$ antibracket defined as:
\[
\{A,B\}_\Lambda\doteq \Lap_\Lambda(AB)-\Lap_\Lambda(A)B-(-1)^{|A|}A\Lap_\Lambda(B)\,.
\]
We can conclude that the scale $\Lambda$ QME is the condition that:
\[
\deL V+\frac{1}{2}\{V,V\}_\Lambda-i\hbar\Lap_\Lambda V=0
\]
This is exactly the form of the regularized {\qme} provided in \cite{Costello}.
\section{Conclusions and outlook}
%
In this paper we presented a formulation of the BV quantization in the framework of perturbative algebraic quantum field theory. Our result is based on the idea to apply to the classical structure of the BV complex the methods of deformation quantization. The quantum BV algebra is equipped with the star product $\star$ and the renormalized time-ordered product $\TR$. We showed that the second of these structures is an associative binary operation on a suitably chosen domain. This result is crucial for incorporating the renormalization into the BV formalism. It is also interesting for the conceptual understanding of perturbative quantum field theory in the algebraic setting. Up to now the renormalized time ordered products were understood only as multilinear operations on the space of local functionals.

In our discussion of the BV quantization we first considered a subalgabra of the BV complex consisting of regular objects. This was important in order to make contact with the standard approach \cite{Batalin:1977pb,Batalin:1981jr}. We showed that the quantum master equation and the quantum BV operator arise naturally in the algebraic setting and the corresponding formulas agree with those used in the path integral formalism. 
The algebraic structure of the quantum algebra is determined by two products: $\star$ and $\T$. The first one is the operator product and the second is equivalent to the pointwise product. They coincide if the arguments are time ordered. The time ordering operator $\TT$ can be used to transport the classical structure into the quantum algebra. We showed that one can define in this way the time-ordered vector fields and the antibracket. The same cannot be done for the $\star$-product, since it is non commutative.
Various relations between the products $\star$ and $\T$ lead to interesting algebraic properties. For example we showed, that the operator $\Lap$, which appears in the {\qme} characterizes the difference between the ideals generated by equations of motion with respect to $\star$ and $\T$, i.e. $i\hbar \Lap=\{.,S_0\}_{\TT}-\{.,S_0\}_\star$.

We provided formulas for the {\qme} and the quantum BV operator that don't involve explicitly the potentially problematic operator  $\Lap$ and can be therefore generalized to the renormalized case:
\begin{align}
0&=\{e_{\sst{\TT}}^{iV/\hbar},S_0\}_\star\,,\label{summary1}\\
\hat{s}X&=e_{\sst{\TT}}^{-i V/\hbar}\T\left(\{e_{\sst{\TT}}^{ iV/\hbar}\T X,S_0\}_{\star}\right)\,.\label{summary2}
\end{align}
This is a completely new result and it gives an algebraic interpretation of the BV quantization.
In the next step we reformulated in the algebraic setting some important results of the BV formalism, known up to now only in the path integral formalism. For example we showed that the interacting field $R_V(F)$ doesn't depend on the gauge fixing if $F$ is in the the kernel of the quantum BV operator $\hat{s}$ and the {\qme} holds. We also provided a simple argument that the cohomologies of $\hat{s}$ obtained for different choices of the gauge fixing are equivalent.

The most important result of this paper concerns the renormalization. We proposed a very natural and straightforward way to define the renormalized counterparts of the {\qme} and the quantum BV operator. We simply replaced the time ordered product $\T$ with the renormalized one $\TR$ in definitions (\ref{summary1}) and (\ref{summary2}). 
In this way we directly obtained well defined expressions and it was nowhere necessary to introduce regularizations in intermediate steps. There remains, of course, the freedom of finite renormalizations governed by the renormalization group.
We showed that both the {\qme} and the quantum BV operator transform correctly under its action. To obtain this result we needed to make one more generalization, namely we had to move to the more abstract level of natural transformations. In this way we formulated the {\qme} and the quantum BV operator in the adiabatic limit.

Another important result of our paper is the use of the anomalous master Ward identity of Brennecke and D\"utsch \cite{BreDue} in order to write the {\qme} and the quantum BV operator in more explicit terms. The first application of the {\mwi} in the context of field-antifield formalism in Yang-Mills is due to S. Hollands \cite{H}. We showed that under the usual assumptions the {\qme} and the quantum BV operator can be written as:
\begin{align*}
0&=\frac{1}{2}\{V+S_0,V+S_0\}_{\TTR}-i\hbar\Lap_r(V)\,,\\
\hat{s}X&=\{X,V+S_0\}_{\TTR}-i\hbar\Lap_V(X)\,.
\end{align*}
The resemblance to the non-renormalized formulas is remarkable. The main feature is the substitution of the divergent, interaction-independent operator $\Lap$ with a well defined, but non-linear map $\Lap_r$, whose derivative induces an interaction-dependent operator $\Lap_V$.

The formalism we propose can be now applied in concrete examples. The case of the Yang-Mills theory was to some extent already treated by S. Hollands in \cite{H}. The formulas we obtained agree with the ones postulated in \cite{H}, but in our framework they arise naturally and are part of a more general setting. Another interesting application would be the treatment of general relativity. The classical theory was already investigated by us in \cite{FR}. Now, with the general quantization scheme at hand we can approach the problem of quantizing gravity, as an effective theory, in the framework of locally covariant quantum field theory, following the program proposed in \cite{F,BF1}. This is currently under investigation.
\section*{Acknowledgements}
We would like to thank R. Brunetti and P. Lauridsen-Ribeiro for enlightening discussions and remarks. One of us (K.F) also profited from the series of seminars focused on the BV formalism, organized during the winter term 2007/08 by The Center for Mathematical Physics Hamburg (ZMP). Furthermore the second author (K.R) wants to thank G. Barnich and J. Zahn for inspiring discussions and comments.
\appendix
%
\section{Appendix}\label{theta:proof}
We present here the proof of relation (\ref{theta0}).
\begin{proposition}\label{brackettheta}
Let $S_0$ be the quadratic term of the action with $\#\af=0$ and $\theta_0$ the free BRST operator. 
Assume that $S_0$ is invariant with respect to the free BRST transformation, i.e.:
\be\label{Stheta}
 \{S_0,\theta_0\}_{\TT}=0\,.
\ee
Then:
\[
\left\{e_{\sst{\TT}}^{i V/\hbar}\T X,\theta_0\right\}_\TT=\left\{e_{\sst{\TT}}^{iV/\hbar}\T X,\theta_0\right\}_\star\,.
\]
\end{proposition}
\begin{proof}
To prove this identity we first note that:
\begin{multline}\label{theta0:exp}
\int\! dx\,\TT\left( \TT^{\minus}\frac{\delta}{\delta\ph(x)}(e_{\sst{\TT}}^{iV/\hbar}\T X)\cdot \TT^{\minus}\theta_0(x)\right)=\\
\int\! dx\, m\circ e^{i\hbar\DDp}\left(\frac{\delta}{\delta\ph(x)}(e_{\sst{\TT}}^{iV/\hbar}\T X)\otimes \theta_0(x)\right)=\\
\int\! dx\frac{\delta}{\delta\ph(x)}(e_{\sst{\TT}}^{iV/\hbar}\T X)\cdot\theta_0(x)+i\hbar\int\! dxdydz\ \Delta_D(y,z) \frac{\delta \theta_0(x)}{\delta\ph(y)}\frac{\delta^2}{\delta\ph(x)\delta\ph(z)}(e_{\sst{\TT}}^{iV/\hbar}\T X)\,.
\end{multline}
Now it remains to prove that 
\be\label{theta0:star}
\int dx\frac{\delta}{\delta\ph(x)}(e_{\sst{\TT}}^{iV/\hbar}\T X)\cdot\theta_0(x)=\int dx \frac{\delta}{\delta\ph(x)}(e_{\sst{\TT}}^{iV/\hbar}\T X)\star\theta_0(x)\,,
\ee
and that the second term of the expansion (\ref{theta0:exp}) vanishes. Actually both results can be obtained in a similar way. We start with the second one. From (\ref{Stheta}) it follows that: 
\[
\int\! dxdydz\ \Delta_D(y,z) \frac{\delta \theta_0(x)}{\delta\ph(y)}\frac{\delta^2 F}{\delta\ph(x)\delta\ph(z)}=
m\circ(\DDp)^2\left(\int\! dx\,\theta_0\frac{\delta S_0}{\delta\ph(x)}\otimes F\right)=0\,,
\]
for an arbitrary  argument $F\in\TT(\BV_\reg(M))$. To show (\ref{theta0:star}) we use a similar reasoning, but this time with the causal propagator:
 \[
\int\! dxdydz\ \Delta(y,z) \frac{\delta \theta_0(x)}{\delta\ph(y)}\frac{\delta^2 F}{\delta\ph(x)\delta\ph(z)}=
m\circ\DDp\circ\DC\left(\int\! dx\,\theta_0\frac{\delta S_0}{\delta\ph(x)}\otimes F\right)=0\,,
\]
It follows now that
\[
\TT\left( \int\! dx\,\TT^{\minus}\frac{\delta}{\delta\ph(x)}(e_{\sst{\TT}}^{iV/\hbar}\T X)\cdot \TT^{\minus}\theta_0(x)\right)=\int dx\frac{\delta}{\delta\ph(x)}(e_{\sst{\TT}}^{iV/\hbar}\T X)\star\theta_0(x)\,,
\]
so to end the proof we need to check
\[
\TT\left(\int\! dx\, \TT^{\minus}\frac{\delta}{\delta \ph^\ddagger(x)}(e_{\sst{\TT}}^{iV/\hbar}\T X)\cdot  \TT^{\minus}\frac{\delta\theta_0}{\delta \ph(x)}\right)=\int\! dx\,\frac{\delta}{\delta \ph^\ddagger(x)}(e_{\sst{\TT}}^{iV/\hbar}\T X)\star \frac{\delta\theta_0}{\delta \ph(x)}\,,
\]
but this is trivially fulfilled, since $\theta_0$ is linear and  hence $\frac{\delta\theta_0}{\delta \ph(x)}$ doesn't depend on fields anymore.
\end{proof}

\end{document}